\documentstyle[12pt]{article}

\advance \topmargin by -\headheight
\advance \topmargin by -\headsep

\evensidemargin \oddsidemargin
\begin{document}
\def\rf#1{(\ref{eq:#1})}  \def\lab#1{\label{eq:#1}}

\def\rf#1{(\ref{eq:#1})}
\def\lab#1{\label{eq:#1}}
\def\foot#1{\footnotemark\footnotetext{#1}}
\def\br{\begin{eqnarray}}           \def\er{\end{eqnarray}}
\def\be{\begin{equation}}           \def\ee{\end{equation}}
\def\({\left(}                      \def\){\right)}
\def\nonu{\nonumber}
\def\al{&&\kern -50pt}
\def\norm{\normalsize\baselineskip=26pt}

\topmargin -.6in
\def\a{\alpha}
\def\b{\beta}                       \def\bt{\tilde \beta}
\def\d{\delta}                      \def\D{\Delta}
\def\eps{\epsilon}
\def\g{\gamma}                      \def\G{\Gamma}
\def\grad{\nabla}
\def\h{{1\over 2}}	\def\i{{\rm i}}  	\def\e{{\rm e}}
\def\k{\kappa}
\def\l{\lambda}                     \def\L{\Lambda}
\def\m{\mu}
\def\n{\nu}
\def\o{\omega}                      \def\O{\Omega}
\def\p{\phi}                        \def\P{\Phi}        \def\vp{\varphi}
\def\pa{\partial}
\def\pr{\prime}
\def\ra{\rightarrow}
\def\s{\sigma}                     
\def\t{\tau}
\def\th{\theta}   \def\Th{\Theta}   \def\vth{\vartheta}  \def\va{\vartheta}
\def\ti{\tilde}
\def\wti{\widetilde}
\def\vareps{\varepsilon}
\def\veps{\varepsilon}
\def\tg{\bigtriangleup}
\newcommand{\ct}[1]{\cite{#1}}
\newcommand{\bi}[1]{\bibitem{#1}}

\begin{titlepage}
\vspace*{-1cm}
\rightline{UBTH-0494}
\rightline{hep-th/9404097}
\vfil
\centerline{\Large\bf  Reality of Complex Affine Toda Solitons}
\vfil
\begin{center}
{\large Z. Zhu   and   D. G. Caldi}\\
\vskip 20pt
{Department of Physics} \\
{State University of New York at Buffalo} \\
{Buffalo, N.Y. \ \ \ 14260-1500}\\
\end{center}
\vfil
\baselineskip=26pt

There are infinitely many topological solitons in any given complex
affine Toda theories and most of them have complex energy density.
When we require the energy density of the solitons to be real, we
find that  the reality condition is related to a simple ``pairing
condition.''  Unfortunately, rather few soliton solutions in these
theories survive the reality constraint,  especially if one also
demands positivity.  The resulting implications for the physical
applicability of these theories are briefly discussed.

\end{titlepage}

\baselineskip=26pt

\section{Introduction}

\setcounter{equation}{0}
Toda field theories (TFT's) [1--14] provide a large class of integrable
two--dimension models which include not only conformal field theories
(CFT's) but also massive deformations away from conformality.  TFT's
are based on an underlying Kac-Moody algebra $g$. If $g$ is a
finite-dimensional, semi-simple Lie algebra, the theory is a CFT.  But
if $g$ is an affine Kac-Moody loop algebra, then the affine Toda field
theory (ATFT) is massive, but still integrable, and these theories are
explicit examples of Zamolodchikov's \ct{zam} integrable massive
deformations of CFT's.  In order to be deformations of {\em unitary}\/
CFT's, it is required [1--4], somewhat paradoxically, that the coupling
be purely imaginary in the ATFT.

In this letter, we are mostly concerned with these complex ATFT's,
i.e., ATFT's with imaginary coupling constant. In
\cite{hollo1,mm,ACFGZ,us}, explicit soliton solutions have been
obtained by using Hirota's method \cite{hir}.  Soliton solutions can
also be found using a vertex operator method \cite{lot,otu1,otu2}.
Complex ATFT's are generalizations of the sine-Gordon theory. However,
unlike sine-Gordon theory, which has only one soliton and one
anti-soliton, here we have infinitely many solitons parameterized by
an integration constant $\xi$.  This is undesirable since it would be
difficult to deal with infinitely many particles of the same mass. To
make matters worse, most of them have complex energy density.  The
problems of infinitely many solitons and complex energy density are
related. One would like a physical principle to select at most one
soliton for each given topological sector, and one hopes that soliton
will have real energy density. The physical principle we are are going
to use in this paper is the reality of the energy density, since
complex energy density defies physical interpretation.  It turns out
that for each topological sector there is at most one ``real''
soliton, and the reality is associated with some ``pairing
condition.''

The Lagrangian density of affine Toda field theory can be written in
the form
\be
  {\cal L}={1\over2}\, \partial_\mu \vp \; \partial^\mu\vp -
  {2\over\psi^2}{m^2\over(\i \b)^2}\sum_{j=0}^r n_j(\e^{\i \b\a_j\cdot\vp}-1)
  = {1\over2}\, \partial_\mu \vp \; \partial^\mu\vp - U(\vp),
  \lab{lag}
\ee
and Hamiltonian density is given by:
\be
  {\cal H} ={1\over2}(\partial_t \vp)^2 +{1\over2}(\partial_x \vp)^2 + U(\vp).
      \lab{Hamiltonian}
\ee
The field $\vp(x,t)$ is an $r$-dimensional vector, $r$ being the rank
of some finite-dimensional semi-simple Lie algebra $g$.  The
$\alpha_j$'s, for $j=1,...,r$, are the simple roots of $g$; $\psi$ is
the highest root: $\psi=\sum_{j=1}^r n_j\a_j$, where the $n_j$'s are
positive integers, and $n_0=1$.  $\a_0=-\psi$ is the extended root of
$\hat g$, the affine extension of $g$.  The coupling constant is $\i
\b$, $m$ is the mass parameter, and $U(\vp)$ is called the Toda potential.
Note, in general repeated indices are not summed over in this paper.

The equations of motion are
\be
  \partial^2 \vp = - {2\over \psi^2} {m^2\over\i \b}\sum_{j=0}^r n_j \alpha_j
  \e^{\i \b \alpha_j\cdot\vp},   \lab{eom}
\ee
where  $\vp$ can be written as
  $$\quad \vp =  \sum_{j=1}^r({2 \a_j/\a_j^2})\, \vp^j. $$

Hirota's method consists of a nonlinear transformation of variables so
that the new equations of motion can be written in Hirota's bilinear
form, and then the equations can be solved perturbatively.  Because of
the integrability, the perturbative series terminates at some finite
order.  The transformation for these theories is:
\be
  \vp^j = - {1\over \i \b} (\ln \t_j - l_j \ln \t_0),  \lab{trans1}
\ee
where $l_j = (\alpha_j^2 / \psi^2) n_j$ forms an integer null vector
of the extended Cartan matrix $K_{jk} =2 \alpha_j \cdot \alpha_k /
\alpha_k^2$.  For each non-zero eigenvalue $\lambda$ of the matrix
$L_{jk} = l_j K_{jk}$, one can find a single soliton solution of the form:
\be
   \tau_j(x, t) = 1 + \sum_{k=1}^{\k l_j} \d_j^{(k)} \e^{k\G}, \lab{solution}
\ee
where $\G = \g (x - v t) + \xi$, $v$ is the velocity of the soliton, and
$\g$ satisfies $ \g^2 (1 - v^2) = m^2 \l $, and is assumed to be
positive for simplicity of discussion.
$\k$ is some integer \cite{ACFGZ} which has possible values 1, 2, and 3.
$\xi$ is an integration constant, and is in general complex.

Since for single-soliton solutions of the form \rf{solution}, all the
fields depend on $x$ and $t$ only through $\G$, we can rewrite the
Hamiltonian as:
\be
  {\cal H} = { m^2\l\,(1+v^2) \over 2(1-v^2)}\;
   \left({{\rm d}\vp \over {\rm d}\G}\right)^2  + U(\vp(\G)).
\ee
Since the Hamiltonian density should transform as a component of a
second-rank tensor under Lorentz transformations, by relating the
Hamiltonian of  a soliton with velocity $v$ to one at rest, we find
  $$ {1\over2} m^2\l\, \left( {{\rm d}\vp \over {\rm d}\G}\right)^2 =
     U(\vp(\G)). $$
In particular, when $v=0$, we have:
\begin{equation}
  {1\over 2} (\partial_x \varphi)^2 = U(\varphi). \lab{KP}
\end{equation}

We will make some simplifying choices of parameters. Since Re($\xi$)
only affects the position of the center of mass of the soliton, it can
be absorbed by a shift in $x$.  Furthermore, we want to study the
effect of Im($\xi$) on topological charge and other properties of the
soliton, so we will take $\xi$ to be purely imaginary: $\xi =\i\;
\theta$.  Since we will only discuss the classical solution, the mass
parameter $m$ and the coupling parameter $\b$ can always be scaled to 1,
$m = \b = 1$.  Finally, because the velocity of a soliton does not
affect the reality of the energy density, we will take $v=0$ whenever
discussing single solitons.

For single-soliton solutions, the $\tau_j$'s can always be factorized to
have simple factors of the form $z_0 = 1 + c\; \e^{\g x + \i\theta}$,
where c is a complex number.  As $x$ goes from $-\infty$ to
$+\infty$,  $z_0$ traces out a straight line on the complex $z$-plane,
starting from $z = 1$. Since $\ln z$ is a multivalued function in the
complex domain, we will take the convention that $\ln z_0(x\rightarrow
-\infty) = 0$. Then as long as the line does not go through the origin,
$\ln z_0$ will behave as a continuous and single-valued function.
$\theta$ will be called singular if and only if $\tau_j$ goes through
the origin of the complex plane for some $j$.  Singular points divide
the unit circle into many continuous sectors.

The topological charge of a soliton is defined as:
\begin{equation}
   T = {\beta \over 2 \pi} \int_{-\infty}^{+\infty}\; \partial_x \varphi(x,t)
     = {\beta \over 2 \pi} \left( \lim_{x \rightarrow \infty} \varphi(x,t)
     - \lim_{x \rightarrow -\infty} \varphi(x,t) \right).
\end{equation}
In the above convention, we have
$\lim\, \varphi\, (x \rightarrow -\infty,\; t) = 0$, so
\begin{equation}
   T_j = {\i \over 2 \pi} \lim_{x\rightarrow \infty} ( \ln \tau_j (x,t)
     - l_j \ln \tau_0 (x,t) ). \nonu
\end{equation}
As long as $\th$ is not singular, all fields will be continuous
functions of $\th$, and the topological charge should be as well,
except that the topological charge can only take discrete values.
Hence all $\theta$ in one continuous sector will have the same
topological charge, and each sector on the unit circle corresponds to
a topological sector of the soliton solutions.  For each given class
of soliton solutions, the number of possible topological charges is no
more than the number of singular points. Explicit topological charges
of $su(n)$ affine TFT's have be given in \cite{mcghee}.

In the following sections we will discuss for various classes of
ATFT's which solitons have real (and positive) energy density.

\section{su(n) and sp(2n)}
\setcounter{equation}{0}
The energy density of   $su(n)$ affine TFT can be written in terms of
$\tau$-functions as:
\begin{equation}
   {\cal H} = \sum_{j=0}^{n-1} \left[ {\tau_j^\prime \over \tau_j}
   \left( {\tau_{j+1}^\prime \over \tau_{j+1}} - {\tau_j^\prime \over \tau_j}
   \right) - {{\dot \tau_j}\over \tau_j} \left( {{\dot \tau_{j+1}}
   \over \tau_{j+1}} -  {{\dot \tau_j}\over \tau_j} \right)
   + \left (1-{\tau_{j+1} \tau_{j-1}  \over \tau_j^2 } \right)
   \right]. \lab{H1}
\end{equation}
For each given integer $1 \leq s \leq n-1$, there is a class of
solutions given by:
\begin{equation}
   \tau_{j}=1 + \e^{2\pi\i sj/n+\Gamma}, \qquad
   \mbox{for $0 \leq j \leq n-1$}, \lab{single1}
\end{equation}
where $\Gamma = \gamma x + \i \theta $, and $\gamma = \sqrt{\lambda_s}
= 2 \sin (s \pi / n)$, and we have assumed zero velocity. It is clear
that $\theta$ is singular if and only if
$$ \theta + {2 \pi s j/n} = \pi\ {\rm mod}\ 2\pi, \quad
   \mbox{for some integer $j$}. $$
Singular points are uniformly spaced on the unit circle and divide
it into equal sectors. Since $\theta = \pi$ is always a singular
point, singular points are symmetrical about the real axis.  For the
$\tau$-functions in \rf{single1}, we have:
\begin{equation}
  \tau_j^\prime = \gamma (\tau_j-1), \qquad
  \tau_{j-1}\tau_{j+1} = \tau_j^2 -\lambda_s (\tau_j-1), \qquad
  \sum_j {1 \over \tau_j \tau_{j+1}} = \sum_j {1 \over \tau_j},\nonu
\end{equation}
so the energy-density function can be simplified to:
\begin{equation}
   {\cal H } = 2 \lambda_s\sum_{j=0}^{n-1} {\tau_j - 1 \over \tau_j^2}.
    \lab{sunH}
\end{equation}

First we will show that the above Hamiltonian density ${\cal H}$ is
real if $\theta$ is singular or is at the mid-point of a sector on the
unit circle.  We accomplish this by demonstrating that for each term
in ${\cal H}$, we can find another term conjugate to it, i.e., for
each $j$, we can find a $j^\prime$ such that
\begin{equation}
  \tau_j = \tau_{j^\prime}^*. \lab{pairing}
\end{equation}
The above equation will be referred to as the ``pairing condition.''
For solitons given in \rf{single1} it is equivalent to $2 \pi s (j +
j^\prime) /n + 2 \theta = 0\ {\rm mod}\ 2 \pi$, or geometrically, the
points $2 \pi s j^\prime /n + \theta$ and $2 \pi s j/n + \theta$
should be symmetrical about the real axis on the unit circle.

Let $\theta$ be a mid-point of some sector. Then for every integer
$j$, ${2 \pi s j/n} + \theta$ is also a mid-point for some sector.
Since the singular points are distributed symmetrically about the real
axis,  the mid-points are also, therefore the pairing condition can
be satisfied if $\theta$ is a singular point or a mid-point. Even
though the singular points give real-energy densities, the singularity
renders them unacceptable. All the mid-points can be written in the
form: $\theta = \pi s (2j+1) /n - \pi$, for some integer $j$. When
$s$ and $n$ are mutually prime, these $\theta$-values coincide with
the examples discussed in \cite{hollo1,Evans}.

Next we will show that these are the only $\theta$-values which will
give real-energy densities.  To prove this, we note that if ${\cal H}$
is real for all $x$, then $\partial_x {\cal H}$ is also real for all
$x$, in particular for $x = 0$. It turns out that the following
equation
\begin{equation}
  {\rm Im}\Bigl( \partial_x {\cal H} (x=0, \theta) \Bigr) = 0, \lab{condition}
\end{equation}
is restrictive enough to select only the mid-points of some continuous
sectors. Since
\begin{equation}
   \partial_x {\cal H} (x,\; \theta) = 2 \lambda_s \sum_{j=0}^{n-1}
    {(2 - \tau_j) \tau_j^\prime  \over \tau_j^3 }, \nonu
\end{equation}
at $x = 0$, $\tau_j(x=0) = 1 + \e^{\i {2 \pi j s /n } + \i \theta}
= 2 \cos({j s \pi \over n} + {\th\over 2}) \exp(\i ({j s \pi \over n}
+ {\th\over 2}))$, equation
\rf{condition} can be simplified to:
\begin{equation}
 {{\rm d} \over {\rm d} \theta} \sum_{j=0}^{n-1} {1 \over \cos^2\({\pi j s/n}
     + {\theta / 2}\) } = 0. \lab{sum}
\end{equation}
To prove that the summation in the above equation has only mid-points
as its minima, we reorganize it to pair up the sectors symmetric about
the real axis.  The partial sum of the two paired sectors can be
written in the form: $ \displaystyle {{1\over \cos^2 (\theta_0 /2 +
\epsilon /2)} + {1\over \cos^2 (\theta_0 /2 - \epsilon /2)}}$, where
$\theta_0$ can be chosen as the mid-point of the sector above the
real axis, and $\epsilon$ is the distance from $\theta$ to the
mid-point of the sector it is in. Because the function $1/\cos^2 x$
is concave in the domain $(0, \pi/2)$, the sum of the pair will
increase monotonically as $\theta$ moves away from a mid-point towards
a closest singular point. One or more pairs will blow up as $\theta$
reaches a singular point, so the mid-point is a minimum point for
every pair, therefor a minimum for the whole sum in \rf{sum}. It is
also obvious that the mid-points are the only non-singular extrema
for the sum.

When $\theta$ is taken to be a mid-point, the energy density is not
only real but also positive definite, which can be seen as follows.
\begin{equation}
   {\cal H} = \lambda_s
   \sum_j {1 + y \cos \theta_j \over (y + \cos \theta_j)^2 },
\end{equation}
where $y = \cosh \gamma x \geq 1$, and the summation runs over all the
mid-points. Since the set of mid-points only depends on the
number of sectors on the unit circle, we only need to do the summation
for the class $s=1$. The positive definiteness can be written down
explicitly as:
\begin{equation}
  \sum_{j=1}^n  {1 - y \cos {(2j-1)\pi\over n} \over
    (y - \cos {(2j-1)\pi \over n})^2 } \geq 0, \quad \mbox {for}\ y \geq 1.
\end{equation}
For $n \leq 10$, we can do the above summation by ``brute force.'' The
summation can be written in the form of a positive constant divided by
a number of positive-definite factors, therefor it is positive
definite itself. With the aid of numerical calculations, we have found
that this pattern persists for $n$'s up to 50. We believe that this
pattern is true for all the $n$'s, but because the number of factors
one has to deal with grows exponentially with $n$, a general
analytical proof has eluded us.  A side conclusion from this is that
the soliton energy density profile is of the form $1/\cosh gx$ rather
than Gaussian for $|x| >>1$.

If $\tau_0$ is real, i.e.\ if $\theta = 0$, then from \rf{trans1} and
$\tau_j^* = \tau_{j^\prime}$, we will have $\varphi_j^* =
-\vp_{j^{\prime}}$, which is the ``twisted reality condition'' in
\ct{hollo1,Evans}.  However, most of the real-energy-density
solutions do not satisfy the twisted reality condition. One simple way
to see this is that there are $n$ vertices in the Dynkin diagram, but
there are only $n-1$ non-trivial components of the $\varphi$ field.
We also point out that the pairing for the $su(n)$ affine Toda
solitons can always be implemented according to some reflection
symmetry of the underlying Dynkin diagram, even though it is not
necessary.  In general, the existence of symmetries of Dynkin diagrams
is neither necessary nor sufficient to generate non-singular
real-energy-density solitons.

{}From the above discussion, we see that the reality of the Hamiltonian
for a single soliton is dependent on choices of the parameter $\th$,
and is not protected by any symmetry principles. Hence it could be
destroyed by perturbations or scattering with other solitons. The
reality of a two-soliton scattering solution can be discussed in a
similar fashion. The two-soliton scattering solutions of $su(n)$
affine Toda solitons are given by:
\begin{equation}
  \tau_j = 1 + \e^{{\i 2\pi s_1 j \over n} + \Gamma_1}
   + \e^{{\i 2\pi s_2 j \over n} + \Gamma_2}
   + A\; \e^{{\i 2\pi (s_1+s_2) j \over n} + \Gamma_1+ \Gamma_2},
\end{equation}
where $A = {\displaystyle {\cosh \omega - \cos {\pi(s_1 -s_2)\over n} \over
\cosh \omega - \cos {\pi(s_1 +s_2)\over n}}}$, and $\omega$ is the rapidity
difference of the two solitons.
  First we impose the pairing
condition on the $\tau$ functions, which can be satisfied if and only if
we can find an integer $j$ such that
\begin{equation}
   \theta_1 = { s_1 (2 j+1) \over n} \pi - \pi \ {\rm mod}\ 2\pi, \qquad
   \theta_2 = { s_2 (2 j+1) \over n} \pi - \pi \ {\rm mod}\ 2\pi, \nonu
\end{equation}
i.e., the scattering solution can be paired if the two solitons can be
paired simultaneously. From \rf{H1}, one sees that the energy density
will be real if the solution can be paired. This observation can also
be generalized to any N-soliton solutions. The necessary part is
harder to prove, since one has to do the summation.  It is enough to
point out that there are explicit examples of scattering solutions
having complex energy density even though each participating soliton
is real.  Furthermore, even when the energy density of the scattering
solution is real, there are cases where it is no longer positive
definite.

Since the soliton solutions of $sp(2n)$ affine TFT can be unfolded
\cite {otfold,mm} to soliton solutions of $su(2n)$ affine TFT, we do
not have to discuss them completely separately. The single-soliton
solutions of $sp(2n)$ are given by:
\begin{equation}
   \tau_j = 1 + 2\cos{\pi s j\over n}\; \e^\Gamma + \cos^2 {\pi s\over 2n}
    \;\e^{2 \Gamma} \quad \mbox{for $0 \leq j \leq n$.} \lab{sp}
\end{equation}
We can unfold each of these to a $su(2n)$ affine Toda soliton by
extending the range of $j$ in the above equation to $0 \leq j \leq
2n$, which gives an $\kappa=2$ soliton of $su(2n)$ affine TFT, and its
energy density is unaffected by the unfolding process.  Since all the
coefficients in \rf{sp} are real, it is clear that the only possible
values of $\theta$ that could satisfy the pairing condition are:
$\theta = 0, \pm \pi/2, \pi$.  For $\theta = 0, \pi$, the solutions
\rf{sp} are always singular.  For $\theta = \pm \pi/2$, the pairing
condition can be simplified to the following question: for an integer
$j_1$, can we find a $j_2$ such that $\cos(\pi s j_1/ n) = - \cos(\pi
s j_2/ n)$?  Let $\displaystyle{{s\over n} = {p\over q}}$, with $p$
and $q$ mutually prime, if $p$ is odd, then the pairing can be done by
letting $j_1 + j_2 = k q$ for some odd integers $k$. But if $q$ is
even, the solution \rf{sp} is singular, so the final conditions for
non-singular pairing are: $\theta = \pm \pi/2$ and $p$ and $q$ are
odd. Going back to \rf{H1}, we can show that the energy density is
real for these cases.  Because the condition of real-energy density is
infinitely over constrained, we believe that there are no other
non-singular real-energy-density soliton solutions. There is strong
numerical evidence supporting this.

\section{so(8)}
\setcounter{equation}{0}

For $\l = 2$, there are three $\kappa=1$ single-soliton solutions:
\be
  \t_0 = \t_1 = 1 + \e^{\G}, \qquad
  \t_3 = \t_4 = 1 - \e^{\G}, \qquad
  \t_2 = 1 +  \e^{2 \G},
\ee
and permutations of indices (1, 3, 4) of the above solution.  All
these three solutions have the same energy density:
\be
   {\cal H} = -16{ 6 + \e^{2\G} + \e^{-2 \G} \over (\e^{2\G} - \e^{-2 \G})^2}.
\ee
Letting $\G = \g x + \i \theta$, where $\gamma = \sqrt{2}$, then:
\be
   {\rm Im}\,({\cal H}) = 16 \sin 2\theta {(\e^{2\g x} - \e^{-2\g x})
     \Bigl( 4\cos^2 \th + 12 (\e^{-2\g x} + \e^{2\g x}) \cos \th
     + (\e^{-2\g x} + \e^{2\g x})^2 \Bigr) \over
     (\e^{4\g x} + \e^{-4\g x} - 2 \cos 2 \theta)^2 }.
\ee
Demanding ${\cal H}$ be real, i.e.\ Im(${\cal H}$) = 0 for all $x$,
then $\sin 2 \theta=0$, or $\theta = 0, \pi, \pm\pi/2$. One can also
easily check that they are also the singular points.

For $\l =6$, the single-soliton solution is given by:
\be
   \t_0 = \t_1 = \t_3 = \t_4 = 1 + \e^{\G}, \quad
   \t_2 = 1 - 4 \e^{\G} + \e^{2\G}.   \nonu
\ee
Since $\vp_2$ is the only non-vanishing component of $\vp$, the
energy density can now be written as:
\be
   {\cal H} = 2 \( \vp_2^{\prime}(x) \)^2
      = -12 \( {-4 \e^{\G} + 2 \e^{2\G} \over 1 - 4 \e^{\G} + \e^{2\G}}
        - {2 \e^{\G} \over 1+ \e^{\G} } \)^2.
\ee
$\theta = 0, \pi$ are the only choices for  reality of ${\cal H}$,
and again  they are also singular points.

We have shown explicitly that there are no non-singular
real-energy-density solitons with $\kappa=1$ in so(8) affine TFT.
This can also be demonstrated by the fact that the singular points are
the only points which can satisfy the pairing condition for these
solutions.  Because of the degeneracy in eigenvalues, we can also have
$\kappa=2$ or~$3$ solitons. It has been shown in \cite{us} that these
high-$\kappa$ solutions are the static limit of scattering solutions
of some $\kappa=1$ solitons.

There are no physical solitons for $g_2$ affine TFT either.

\section{so(2r)}
\setcounter{equation}{0}
For so(2r) affine TFT, the single-soliton energy density can be
written  in terms of tau-functions as:
\begin{eqnarray}
   {\cal H} &=&{1\over 2}\sum_{j,k=1}^r K_{jk} \varphi_j^{\prime}
     \varphi_k^{\prime}+U(\varphi)
     \; = \sum_{j,k=1}^r K_{jk} \varphi_j^{\prime}\varphi_k^{\prime} \cr
     &=&  2 \sum_{j=2}^{r-2} {\varphi_j^{\prime}}^2
     - 2 \sum_{j=2}^{r-3} \varphi_j^{\prime} \varphi_{j+1}^{\prime}
     + 2\( {\varphi_1^{\prime}}^2 + {\varphi_r^{\prime}}^2
     + {\varphi_{r-1}^{\prime}}^2  -  \varphi_1^{\prime} \varphi_2^{\prime}
     -  \varphi_r^{\prime} \varphi_{r-2}^{\prime}
     -  \varphi_{r-1}^{\prime} \varphi_{r-2}^{\prime} \) \cr
    &=& - 2 \sum_{j=0}^{r}{ {\tau_j^\prime}^2 \over \tau_j^2}
     + 2 \sum_{j=2}^{r-3} { \tau_j^\prime \tau_{j+1}^\prime
     \over \tau_j\t_{j+1}} + 2 \( {\tau_0^\prime \over \tau_0}
     + {\tau_1^\prime \over \tau_1} \) {\tau_2^\prime \over \tau_2}
     + 2 \( {\tau_r^\prime \over \tau_r}
     + {\tau_{r-1}^\prime \over \tau_{r-1}} \)
     {\tau_{r-2}^\prime \over \tau_{r-2}}.
\end{eqnarray}

For $\lambda = 2$, there are two independent soliton solutions:
\begin{eqnarray}
   &&\tau_0 = 1 + \e^\Gamma, \quad \tau_1 = 1 - \e^\Gamma,
     \quad \tau_r = 1 \pm (-1)^{r\over 2}\e^\Gamma, \quad
     \tau_{r-1} = 1 \mp (-1)^{r\over 2}\e^\Gamma, \cr
   && \tau_j = 1 + (-1)^j \e^{2\G} \quad (2\leq j\leq n-2).
\end{eqnarray}
By applying the pairing condition, we see that: when $r$ is even,
$\theta = 0,\ \pi,\ \pm \pi/2$; when $r$ is odd, $\theta = 0,\ \pi,\
\pm \pi/2,\ \pm \pi/4,\ \pm 3\pi/4$, are the only solutions, and
$\theta = 0,\ \pi,\ \pm \pi/2$ are always singular.  By evaluating the
energy density explicitly, we can see that these are sufficient and
necessary conditions for the solitons to have real-energy density.

For $\lambda_s = 8\sin^2 \vartheta_s$, where
$\vartheta_s = {\displaystyle{s \pi\over 2(r-1)}}$, \
$(1 \leq s\leq r-2)$,  the solutions are given as:
\begin{eqnarray}
   && \tau_0 = \tau_1 = 1 + \e^\Gamma, \quad
      \tau_r = \tau_{r-1} = 1 + (-1)^s \e^\Gamma, \cr
   && \tau_j = 1 + {2\cos((2j-1)\vartheta_s)\over \cos \vartheta_s}
    \e^\Gamma + \e^{2 \Gamma}, \quad \mbox{for}\ 2 \leq j \leq r-2.
\end{eqnarray}
Since all coefficients in the above equation are explicitly real, the
only possible solutions to the pairing conditions are $\theta = 0,\
\pi,\ \pm \pi/2$.  $\theta$ = 0 or $\pi$ obviously give real energy
densities since all the $\tau$-functions are real, therefore every
term in ${\cal H}$ is real; however, they are also always singular.
For $\theta = \pm \pi/2$, the pairing condition simplifies to the
following question: for a given integer $j_1$, can we find an integer
$j_2$ such that
$\cos (2j_1 - 1) \vartheta_s = - \cos (2j_2 - 1) \vartheta_s$?
Let ${\displaystyle{s \over r-1} = {p \over q}}$, where $p$ and $q$
are mutually prime, then the pairing  can be done if and
only if p is odd, by $j_1 + j_2 - 1 = k q$ for some odd integer $k$.
The energy density is singular if q is also odd.

Now let us argue that these conditions are sufficient and necessary.
In the case when $s$ and $r-1$ are mutually prime, the terms can be
simply paired up as complex conjugate by the the pairing condition,
therefore the energy density is real. If $s$ and $r-1$ are not
mutually prime, let $s=n p$, and $r-1 = n q$, then the energy density
can be reduced to $n$ copies of the energy density of a class $p$
soliton in $so(2q+2)$ affine TFT.  Thus we have proved the
sufficiency. Necessity of the condition is harder to prove directly,
so we turn to numerical verification.

Since the energy density is a continuous function of $\theta$ for most
$x$, let us pick five $x$ values, and see which $\theta$ value can
give real energy density for all five points simultaneously. We
searched all the classes of solitons up to $r \leq 50$. We checked our
numerical calculations against the analytic equation \rf{KP}, and the
total energy (mass) gives the expected value. All $\theta$-values
which satisfy the pairing conditions do give real energy density, and
no new $\theta$ values showed up.  Given that the system is infinitely
over constrained, this should not be a surprise.

The reality of the energy density for multi-soliton solutions \cite{us}
can be similarly discussed.  One can see that the reality here is also
related to the pairing condition.

\section {e-series}
The soliton solutions of $e_6$, $e_7$ and $e_8$ complex affine Toda
theories have been tabulated in \cite{mm} and will not be given here.

There are six $\kappa = 1$ solitons for the $e_6$ affine TFT.  Let us
apply the pairing condition to these solitons.  For the solitons
with eigenvalue $\lambda = 2 (3 \pm \sqrt{3})$, all the
$\delta$--coefficients are real, and the only $\theta$--values are $0$ or
$\pi$, both values being singular.  For $\lambda=3-\sqrt{3}$ solitons,
the possible $\theta$-values are $0, \pi, \pm\pi/3, \pm 2\pi/3$. It is
easy to verify that $\pi$ and $\pm\pi/3$ are singular.  For $\lambda = 3 +
\sqrt{3}$ solitons, the pairing condition gives  the same set of possible
$\theta$-values as for $\lambda=3-\sqrt{3}$ solitons; however, in
this case they are all singular.  Numerical analysis indicates that the
energy density is real if and only if the pairing condition is
satisfied, and the energy density is not positive definite even for
the non--singular $\theta$--values.

For the $e_7$ affine TFT there are seven soliton solutions.  For all
these solitons, $\theta = 0, \pi$ gives real but singular energy
density. The pairing condition can also be satisfied when $\theta =
\pm \pi/2$ for $\lambda = 8 \sin^2 (\pi/9)$, $8 \sin^2 (2\pi/9)$, $8
\sin^2 (4\pi/9)$ solitons. Only the $\lambda = 8 \sin^2 (\pi/9)$ soliton
gives non-singular and positive--definite energy density. There are no
other real--energy--density solitons, at least from the evidence of
numerical studies.

For the $e_8$ affine TFT, there are eight soliton solutions, and for
all of them, $\theta =0, \pi$ satisfies the pairing condition and
gives real energy--density; however, they are all singular.  Numerical
analysis again indicates that there are no non--singular,
real--energy--density solitons in this theory.

\section{Conclusions}

We studied the question of whether all mathematical solutions of
complex affine Toda theories are physically acceptable, and discussed
the consequences of demanding real-energy density for these soliton
solutions.

Because there is this problem that the energy density is in general
complex, one may consider other alternatives besides imposing reality
as a condition: we may simply ignore the imaginary part, or take the
norm of the complex energy density, or ``improve'' the energy density
such that it becomes real for all the solutions.  Each alternative has
its shortfalls: the real part of the energy density is not necessarily
positive definite, the norm of the complex energy density is not
conserved, and since the Lagrangian is intrinsically complex, the
``improving'' approach looks impossible at this time. The reality
condition is in some sense too strong since many topological sectors
have no such solitons, yet it is not strong enough since the energy
density may still be negative.  Furthermore, scattering of two
``real'' solitons may destroy the reality of their combined energy
density.  Even though the energy density for solitons are in general
complex, the total energy is always positive and finite, except when
the energy-density function is singular.

We have focused our attention on the $\kappa=1$ solitons of affine
TFT's based on simply-laced Lie algebras, since higher-$\kappa$
solitons for the simply-laced affine TFT's can be taken as the static
limit of some $\kappa=1$ scattering solutions, and solitons of
non-simply-laced affine TFT's can be unfolded to solitons of their
simply-laced counterparts. We have shown case by case that the
pairing condition is sufficient to guarantee the reality of the energy
density; we also presented strong evidence that it is also necessary.

In conclusion, we have found that the existence of real-energy-density
solitons in complex affine Toda field theories is highly non-trivial,
and is not related to symmetries of the underlying Dynkin diagrams.
Since the Hamiltonian is intrinsically complex, we feel that the
``on-shell'' reality of the energy density for the classical solitons
is a minimum requirement.  Since the reality of the single solitons is
not protected by any principle but imposed as an {\it ad hoc} condition, it
could be destroyed by scattering or perturbation.  In our opinion this
problem deserves further investigation before any consistent
quantization can be meaningfully discussed.


\end{document}